\documentclass{iopjournal}

\begin{document}

\articletype{Paper} 

\title{Mirror Dual Symmetry in Physics}

\author{Lucas Lamata$^1$\orcid{0000-0002-9504-8685}}

\affil{$^1$Departamento de F\'isica At\'omica, Molecular y Nuclear, Facultad de F\'isica, Universidad de Sevilla, Apartado 1065, E-41080 Sevilla, Spain}

\email{llamata@us.es}

\keywords{Quantum Rabi model, Dirac equation, Dirac sea, CPT theorem, antiparticle}

\begin{abstract}
The quantum Rabi model [1] has been a useful and pedagogical quantum model in the past decades, sufficiently simple to be solved analytically and intuitively understood, while sufficiently complex as to provide highly non-trivial eigenstates [2] and a practical description of quantum optical platforms for quantum technologies. The Dirac equation~[3], especially when restricted to 1+1 dimensions, is a simple toy model as well, but its easy diagonalization enabled historically to connect the electron spin to the fermionic statistics, among others. Both models share a symmetry at the purely mathematical level, namely, the spectra of each one has a dual equivalent under energy sign change, that I name a mirror dual symmetry. Usually, one quantizes these equations by assuming a ground state energy for the bosonic mode. But there is another option for the interpretation of the Hamiltonian, as I will argue, that is to assume a total symmetry principle, namely, that the total energy is zero at all times, for either the quantum Rabi model or the Dirac equation, and impose the constraint that every positive energy excitation has a mirror excitation of negative energy. This possibility, which was, apparently, ignored in the times when Paul Dirac was studying the implications of his equation, would avoid the worries in the scientific community that the negative energy solutions would decay until minus infinity, thus obviating the necessity to build a highly artificial Dirac sea, and instead impose what has always been successful in Physics, which is the enforcement of symmetry principles. Assuming a total symmetry principle, many of the problems of current Physics, such as renormalization of quantum gravity, dark matter, and dark energy, may possibly be automatically solved. One obvious result would be the automatic cancellation of the zero point energy, which is an open puzzle in Physics currently, with a discrepancy of 120 orders of magnitude in the theory and experimental measurements. Renormalization of quantum gauge theories, quantum gravity as well, would be automatic at the physical level, without any artificial need of ``sweeping infinities under the carpet''.
\end{abstract}

\section{The mirror dual symmetry}

In this article I introduce a physical principle which, in my view, might be the most fundamental one when trying to elucidate the connection among all physical theories, which I name the ``total symmetry principle''. It basically states that the Universe, as seen globally, has no structure or dimensions, and can be seen, in some sense, as a Hilbert space of dimension 1, or a metric space of dimension 0, to give just two examples. This is partially motivated by arguments by Wheeler and DeWitt [4], regarding a possible dynamical equation for the Universe with single zero energy eigenvalue. Everything that we perceive in our physical reality is just produced inside the Universe, and relative to our particular place, time, and motion inside the Universe [5]. Both the Schr\"odinger equation and Einstein field equation are local, and in my view, physical reality is just created locally, with respect to a given local observer. In this sense, each local observer is the center of his/her/its own local Universe, locally created, which may be totally different from other observers local Universes, as far as they are not sufficiently correlated at the quantum level. The only absolute, according to this hypothesis, would be the total symmetry principle, according to which the Universe has no structure as considered as a whole, only from inside of it.

A way of satisfying the total symmetry principle is just by the concept of the mirror dual symmetry. By construction, the Universe has zero total energy at all times, as well as zero total linear and angular momentum, and zero total charges of any kind of fundamental forces, as far as one enforces this mirror dual symmetry. Namely, that each quantum excitation of, say, energy $E$, momentum $P$, and charges $Q$, has a dual excitation produced at the same point of spacetime, of negative energy, $-E$, negative momentum, $-P$, and negative charges, $-Q$, and this takes place in a mirror dual Universe with sign-changed metric. This is formally equivalent to assume that our Universe has a CPT-symmetric Universe which is dual to ours at every spacetime point, and for every particle. Models of CPT-symmetric Universes have been considered before, although in other contexts [6]. As CPT symmetry is known to hold in our known Physics laws, by construction the possible existence of the mirror dual Universe is trivially allowed, at least mathematically. If one takes the mirror dual symmetry at its most radical consequences, it would be a fundamental symmetry of every physical system, which is more easily seen in the quantum Rabi and Dirac equations, where one could, in a simplified approach, just substitute a creation operator of a bosonic excitation with positive energy, by a dual operator, that creates the positive energy quantum, and a dual negative energy one:

\begin{equation}
a_k^\dag|0\rangle=|1_k\rangle=|E_k\rangle \rightarrow a_k^\dag (\bar{a}_{\bar{k}})^\dag |0\rangle=|1_k \rangle |1_{\bar{k}}\rangle=|E_k\rangle |\bar{E}_{\bar{k}}\rangle
\end{equation}

In the previous equation, the barred magnitudes are the sign-changed ones with respect to the original ones, where $E$ is the energy of the state and $k$ any quantum numbers such as momentum, spin, etc. The total energy is always zero, as well as the momentum, electric charge, etc. In my view, this way of approaching models such as the quantum Rabi and the Dirac equation is just more natural, as it does not artificially single out the positive energy Hilbert space as preferential with respect to the negative energy one. This would be similar as considering Spain as superior to New Zealand, as New Zealand is ``backwards'' in the Earth, with respect to Spain. But from their own perspective, the Spaniards are down, such that it does not seem to be an absolute thing. In my view, the symmetry in simple physical models such as these two is more natural if one just fixes a global total symmetry constraint, and allows for excitations that enforce the mirror dual symmetry.

Another place where there is evidence that this symmetry could be a fundamental tenet of our Universe is when analyzing black-hole radiation [7]. As Stephen Hawking discovered, Einstein theory of gravity, combined with quantum field theory, would have as a consequence the fact that black holes would radiate energy. How? a particle-antiparticle pair would be created near the horizon, satisfying, precisely, the total symmetry constraint, namely, the antiparticle would have negative energy, momentum, spin, charges, etc., with respect to the other one, and it would be at the inner side of the black-hole horizon, with a sign-changed metric. Therefore, it is apparent that also in this example the total symmetry principle is automatically verified, and the mirror dual Universe would be nothing but the inner side of the black hole, beyond the event horizon.

This also points to the fact that our own Universe may be nothing but an holographic projection of a lower dimensional, purely quantum Universe [8], and that gravity may be nothing but a residual entanglement from our mirror dual Universe, when discarding it, what in purely quantum terms is usually named as tracing over the environment. Dark matter may be an effect of a residual entanglement with the mirror dual Universe at galactic scales, and dark energy the corresponding one when considering the whole observable Universe, where the mirror dual Universe would lay beyond the cosmological horizon, in this case. This kind of approach to gravity is also consistent with entropic gravity theories [9], with CPT dual Universes as mentioned above [6], as well as with phenomenological decoherence models that go as the mass square [10], which, in our particular approach to gravity, appears naturally in a master equation when tracing out the mirror dual system of mass $m$  to the same mass system one is analyzing, for mesoscopic to human-size systems, assuming the macroscopic mass fluctuates in a time shorter than the one employed for the calculation of the master equation.

The existence of the mirror dual symmetry and Universe would not just be an artifact from the physical equations with no physical implications, but on the contrary, this could be verified in galactic and cosmological levels, for aiming at explaining dark matter and dark energy with minimal extra assumptions. At human and smaller scales, the known physical laws would not be modified at an effective level, given that gravity is well described by Einstein theory and quantum systems well described at microscopic levels by the Schr\"odinger equation, but experiments of gravity-induced decoherence with nanomechanical resonators could, at some point, be able to falsify (or otherwise) this kind of proposal.

Max Planck said, famously, that he introduced his quantum of energy as a ``desperate act'', given that this was the only seemingly plausible way to explain the experimental data. In my view, the assumption of a total symmetry principle in our Universe, enforced by a mirror dual symmetric Universe, may be my own desperate way to make sense of many inconsistencies in our Physical theories. Only future experimental work may really shed some light as if this may make any real sense, after all.

%
%






\section*{References}

[1] I. I. Rabi, J. R. Zacharias, S. Millman, and P. Kusch, A New Method of Measuring Nuclear Magnetic Moment, Phys. Rev. 53, 318 (1938).\\

[2] D. Braak, Integrability of the Rabi Model, Phys. Rev. Lett. 107, 100401 (2011).

[3] P. A. M. Dirac, The Quantum Theory of the Electron, Proc. Roy. Soc. London A, 117, 610 (1928).

[4] Bryce S. DeWitt, Quantum Theory of Gravity. I. The Canonical Theory, Phys. Rev. 160, 1113 (1967).

[5] Carlo Rovelli, Relational Quantum Mechanics, Int. J. of Theor. Phys. 35 (1996) 1637.

[6] Latham Boyle, Kieran Finn, Neil Turok, CPT-Symmetric Universe, Phys. Rev. Lett. 121, 251301 (2018).

[7] S. W. Hawking, Black holes and thermodynamics, Phys. Rev. D 13, 191 (1976).

[8] G. 't Hooft, Dimensional Reduction in Quantum Gravity, arXiv:gr-qc/9310026

[9] Erik Verlinde, On the origin of gravity and the laws of Newton, JHEP 2011, 29 (2011).

[10] L. Di\'osi, Models for universal reduction of macroscopic quantum fluctuations, Phys. Rev. A 40, 1165  (1989).

\end{document}